\documentclass[aps,prl,twocolumn,superscriptaddress,amsfont,graphicx,nofootinbib,preprintnumbers]{revtex4}%
\usepackage{color,graphicx,epsfig}
\usepackage{ifpdf}
\usepackage{amsmath}
\usepackage{bm}
\usepackage{color}
\usepackage[english]{babel}
\usepackage{graphicx}%
\usepackage{amsfonts}%
\usepackage{amssymb}
\usepackage{braket}
\usepackage{hyperref}
\usepackage{enumerate}

\definecolor{nicered}{rgb}{0.7,0.1,0.1}
\definecolor{nicegreen}{rgb}{0.1,0.5,0.1}
\hypersetup{colorlinks,citecolor= nicegreen,linkcolor= nicered}

\newcommand{\beq}{\begin{equation}}
\newcommand{\eeq}{\end{equation}}
\newcommand{\bea}{\begin{eqnarray}}
\newcommand{\eea}{\end{eqnarray}}

\definecolor{Red}{rgb}{1.,0.,0.}

\def\OMIT#1{}

\begin{document}

\def\Fermilab{Fermilab, P.O.Box 500, Batavia, IL 60510, USA}
\def\Durham{Institute for Particle Physics Phenomenology, Department of Physics, \\ University of Durham, Durham, DH1 3LE, UK}
\def\Buffalo{Department of Physics, University at Buffalo \\ The State University of New York, Buffalo 14260 USA}

\preprint{
\noindent IPPP/16/115, FERMILAB-PUB-16-585-T}
\flushbottom

\title{Direct photon production at next-to-next-to-leading order}

\author{John M. Campbell}     
\email[Electronic address:]{johnmc@fnal.gov}
\affiliation{\Fermilab}

\author{R. Keith Ellis}     
\email[Electronic address:]{keith.ellis@durham.ac.uk}
\affiliation{\Durham}

\author{Ciaran Williams}     
\email[Electronic address:]{ciaranwi@buffalo.edu}
\affiliation{\Buffalo}

\date{\today}
\begin{abstract}
We present the first calculation of direct photon production at next-to-next-to leading order (NNLO) accuracy in QCD.  
For this process, although the final state cuts mandate only the presence of a single electroweak boson, the underlying kinematics resembles 
that of a generic vector boson plus jet topology. In order to regulate the infrared singularities present at this order we use the $N$-jettiness slicing procedure,
applied for the first time to a final state that at Born level includes colored partons but no required jet.  We compare our predictions to 
ATLAS 8 TeV data and find that the inclusion of the NNLO terms in the perturbative expansion, supplemented by electroweak corrections, provides an excellent
description of the data with greatly reduced theoretical uncertainties.
\end{abstract}

\maketitle

\section{Introduction} \label{sec:intro}

Direct (or inclusive) photon production at hadron colliders provides an excellent testing ground for
probing the predictions of the Standard Model (SM) in fine detail.
The LHC, which is currently in its second major data-taking period, provides a powerful tool to study this process~\cite{Khachatryan:2010fm,Aad:2013zba,Chatrchyan:2013mwa,Aad:2016xcr}. 
For the first time, the experimental uncertainties are under such good control that, over a large region of phase space, 
they are significantly smaller than the corresponding theoretical ones. Additionally,
the most recent data from the LHC highlight the fact that existing theoretical tools are inadequate for describing the experimental measurements~\cite{Aad:2016xcr}. 

This remarkable achievement of experimental science challenges the theoretical community to provide more sophisticated predictions 
that have theoretical errors commensurate with the errors in the data. Given the special nature of this final state, the poor description provided by existing
theoretical predictions for this channel has serious ramifications for the LHC program.  Direct photon production ($pp\rightarrow \gamma  + X$),
and the associated process in which a jet is explicitly reconstructed ($pp\rightarrow \gamma  + j$), are the highest-rate electroweak
processes at the LHC.  As such they represent critical standard candles for the exploration
of the SM at the LHC. 
For instance, measurements over a wide range of kinematic configurations -- corresponding to different rapidities and transverse momenta of
the photon -- could be used to provide a precision probe of parton distribution functions (pdfs)~\cite{Carminati:2012mm}.  However, up to now, the large theoretical uncertainty
has meant that this data is not routinely used in fits.
Moreover, the similarity of this process to $Z$+jet
production can also be exploited to provide a better understanding of the $Z(\to \nu\bar\nu)$+jet process, which gives rise to leading backgrounds
in searches for dark matter and supersymmetry.  This is especially useful in the high transverse momentum region,
where there is limited experimental data from the process $pp\rightarrow Z(\to \ell^+\ell^-)+X$~\cite{Bern:2011pa}.
 
Over the last 15 years the theoretical benchmark for direct photon studies has been the next-to-leading order (NLO) Monte Carlo code {\tt JETPHOX}~\cite{Catani:2002ny}.
Recent calculations, implemented in the code {\tt PeTeR}, have extended the NLO prediction to include 
both threshold resummation at the next-to-next-to-next-to leading logarithmic accuracy (N$^3$LL) and 
electroweak Sudakov corrections at leading logarithmic accuracy~\cite{Schwartz:2016olw}. By including the resummed terms the agreement with data is somewhat improved, compared 
to the pure NLO prediction.

It is clearly highly desirable to have a next-to-next-to-leading order (NNLO) prediction for direct photon production that 
can be compared with LHC data. This is the primary aim of this paper. Although direct photon production can be defined merely through fiducial cuts on the photon,
it proceeds at LO in perturbation theory through the recoil of the photon against a quark or gluon. Therefore 
the underlying structure of the calculation is almost identical to that of the $\gamma +j$ process. The presence of a final-state colored parton
means that a NNLO calculation of this process represents a considerable theoretical challenge.

Over the last few years significant progress has been made in the field of NNLO calculations, allowing for the calculation 
of processes involving one~\cite{Chen:2014gva,Ridder:2015dxa,Boughezal:2015aha,Boughezal:2015dva,Boughezal:2015dra,Boughezal:2015ded,Ridder:2016nkl}
or two~\cite{Currie:2013dwa,Currie:2016bfm} massless partons in the final state for the first time. 
One of the developments that has proven very fruitful for computing NNLO corrections is a novel method for regulating the infrared (IR) singular structure known
as $N$-jettiness slicing (or subtraction). Originally used in a NNLO calculation of top quark
decay~\cite{Gao:2012ja}, the method has since been extended and applied to general LHC processes~\cite{Boughezal:2015dva,Gaunt:2015pea}. This method splits 
the phase space into two components based on the global event shape, $N$-jettiness $(\tau_N)$~\cite{Stewart:2010tn}. Crucially, for an $N$-jet final state,
the double-unresolved IR poles occur when $\tau_N =0$, so that $\tau_N > \tau_N^{\rm{cut}}$ corresponds to a region in which the calculation has at
most single-unresolved limits and therefore resembles a NLO calculation.  Furthermore, the cross section in the region $\tau_N < \tau_N^{\rm{cut}}$
can be obtained from a factorization formula derived from soft-collinear effective theory
(SCET)~\cite{Bauer:2000ew,Bauer:2000yr,Bauer:2001ct,Bauer:2001yt,Bauer:2002nz}.
The method can therefore be used as a slicing procedure, with the usual caveat that $\tau_N^{\rm{cut}}$ should be taken as small possible to minimize power corrections
to the below-cut factorization theorem.\footnote{For recent work on reducing the dependence on the power corrections, see refs.~\cite{Moult:2016fqy,Boughezal:2016zws}.}

\section{Calculation} \label{sec:Calc}

In this section we briefly outline the technical details relating to our calculation. In the $N$-jettiness slicing approach the 
calculation naturally splits into two components, corresponding to the contributions above and below $\tau_N = \tau_N^{\rm{cut}}$. 
Below $\tau_N^{\rm{cut}}$, the factorization theorem 
of SCET describes the cross section as a convolution of a process-dependent hard function, $\mathcal{H}$, with process-independent beam functions $\mathcal{B}$ that 
describe initial-state collinear singularities, jet functions $\mathcal{J}$ for final-state collinear singularities, and a soft function $\mathcal{S}$ that 
describes soft radiation. Expansions accurate to $\mathcal{O}(\alpha^2_s)$ that are relevant for our calculation can be found in refs.~\cite{Gaunt:2014xga,Gaunt:2014cfa},
~\cite{Becher:2006qw,Becher:2010pd} and~\cite{Boughezal:2015eha}
for the beam, jet and soft-functions respectively.
The process-independent functions have already been used in the calculation of NNLO corrections to the similar
$W+j$~\cite{Boughezal:2015dva} and $Z+j$~\cite{Boughezal:2015ded} processes.
For our purposes we use the implementation of these contributions in MCFM as outlined in ref.~\cite{Boughezal:2015ded} and also, for the color-singlet case,
in ref.~\cite{Boughezal:2016wmq}.  We have calculated the process-dependent hard function $\mathcal{H}$ using the results for the double-virtual
$pp\rightarrow \gamma j$ matrix elements, calculated in ref.~\cite{Anastasiou:2002zn}. The helicity amplitudes for the $gg\rightarrow \gamma g$ one-loop
contribution, that also enter the hard function, have been computed using analytic unitarity techniques~\cite{Britto:2004nc,Mastrolia:2009dr,Badger:2008cm}.
We have checked that we find agreement between the $N$-jettiness
slicing method and a more traditional Catani-Seymour dipole~\cite{Catani:1996vz} calculation at NLO. 

In addition to the ingredients required from the SCET factorization theorem we also require the pieces associated with $\tau_N > \tau_N^{\rm{cut}}$, 
which corresponds to the NLO calculation of $\gamma +2j$. This process has been studied at NLO in ref.~\cite{Bern:2011pa}. We have re-computed the virtual corrections
to this process using unitarity methods and checked our calculation using an in-house implementation of
the $D$-dimensional numerical algorithm described in ref.~\cite{Ellis:2008ir}.  The amplitudes have also been cross-checked numerically at specific phase space points
using Madgraph5\_aMC@NLO~\cite{Alwall:2014hca}. 
\begin{center}
\begin{figure}[ht]
\includegraphics[width=8cm]{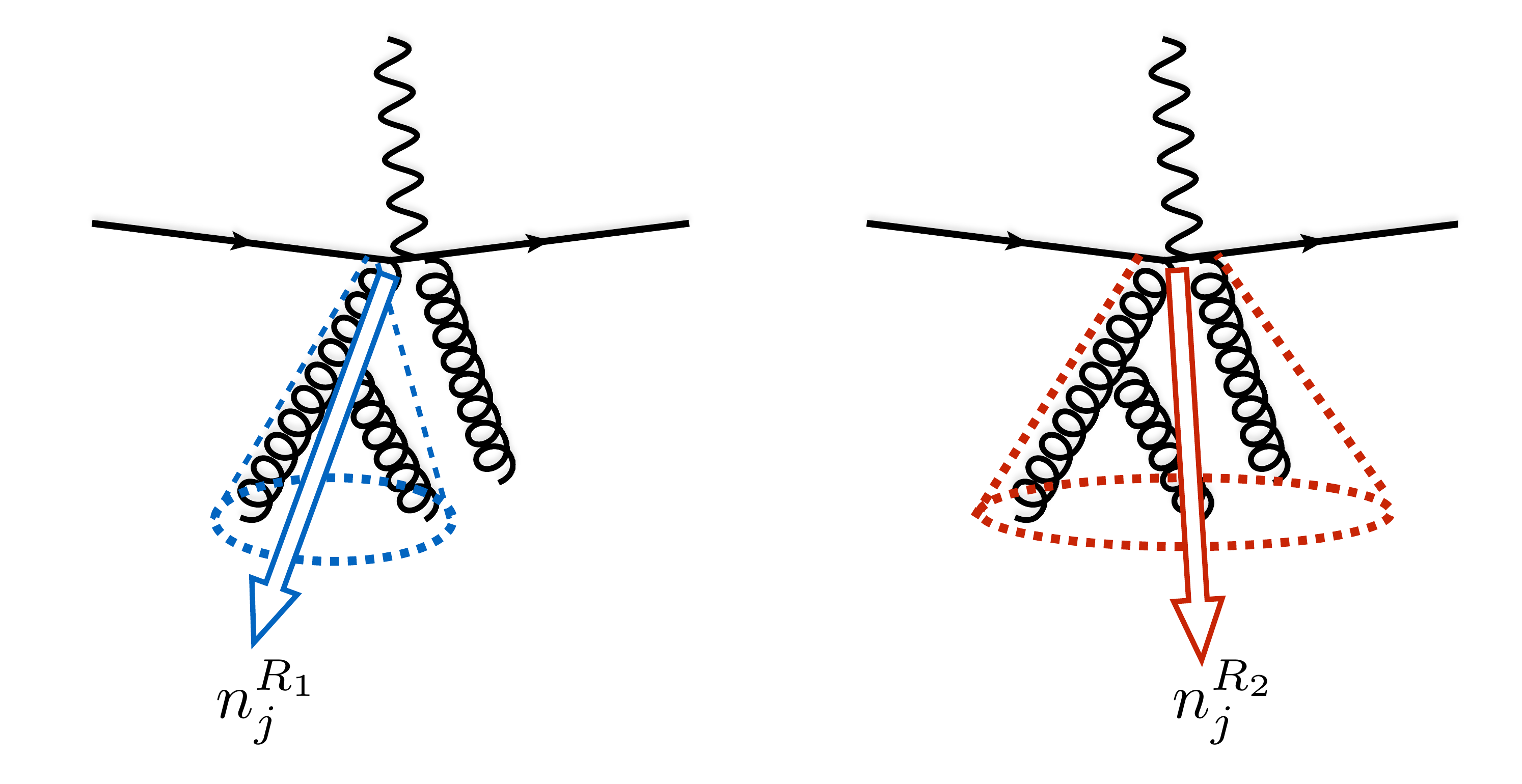}
\caption{An illustration of the dependence on $R$ in the $N$-jettiness algorithm. The same event, clustered with two different $R$ values,
results in differing orientations of the jettiness axis. This results in different power corrections for $\tau_1 > 0$.
In the limit $\tau_1 \rightarrow 0$ the dependence on $R$ vanishes. }
\label{fig:Rtaudep}
\end{figure}
\end{center}

We now turn our attention to the discussion of $N$-jettiness slicing for the case at hand.
Direct photon production is representative of an interesting class of processes to compute at NNLO, where no final-state jet is required but the non-zero recoil of the
photon mandates some colored radiation in the final state.\footnote{Similar studies for inclusive Higgs and $Z$ boson production at finite $p_T$ can be found in refs.~\cite{Chen:2016zka,Ridder:2016nkl,Gehrmann-DeRidder:2016jns}.}
The $N$-jettiness slicing procedure therefore has to be defined in this
context. Clearly the final state parton will induce singularities at NNLO that cannot be regulated by a cut on 0-jettiness, $\tau_0$
(for instance, corresponding to the triple-collinear splitting of a final-state parton). Thus it is clear that a cut must be made on the 1-jettiness event shape
variable, which naturally requires a definition of a jet direction $n_j$. Therefore one has to be careful that the regulating
variable $\tau^{\rm{cut}}_1$, which requires a jet definition, does not interfere with the inclusive nature of the final state, which does not. 

In order to achieve this, we start with the usual definition of $\tau_1^{\rm{cut}}$,
\begin{eqnarray}
\label{eq:tau1def}
\tau_1^{\rm{cut}} = \sum_{k=1}^{M}\min_{i=a,b,1}\left\{\frac{ 2 q_i \cdot p_k}{Q_i}\right\} \;.
\end{eqnarray}
This equation involves the momenta of the parton-level configuration, $\{p_k\}$ and the set of momenta that is obtained
after application of a jet-clustering algorithm, $\{q_i\}$.  The scale $Q_i$ is a measure of the jet or beam hardness, which we
take as $Q_i = 2E_i$.  In order to be well-defined, the contribution from the jet direction that enters in Eq.~(\ref{eq:tau1def})
must correspond to a sufficiently hard jet.  This is guaranteed by the cut on the photon transverse momentum $(p_T)$.
In the Born phase space the transverse momentum of the jet clearly balances that of the photon.  In the real-virtual phase space
this constraint is somewhat relaxed, so that the transverse momentum of the leading jet is constrained by
$p_T^1 > p_T^{\gamma}/2$.  In the double-real contribution the constraint is $p_T^1 > p_T^{\gamma}/3$.  Thus, as long as we
consider sufficiently hard photons, $\tau_1^{\rm{cut}}$ is well-defined.

A subtlety to this procedure still arises in practice.  Although the jet-clustering procedure is only used to identify the jet
direction, with no $p_T$ cut necessary,   there is still a dependence on the cone size $R$. An example of this dependence is
illustrated in Fig.~\ref{fig:Rtaudep}, which makes it clear that, depending on the cone-size, radiation may or may not be
clustered together to form a jet. In the figure the smaller cone $R_1$ results in a different jettiness direction than the
larger cone $R_2$. Crucially, although these two jettiness directions $n_j^{R_1}$ and $n_j^{R_2}$ will differ at large $\tau_1^{\rm{cut}}$,
in the limit in which $\tau_1^{\rm{cut}} \rightarrow 0$, the difference vanishes. Different choices of $R$ will therefore result in
different power corrections at large $\tau_1^{\rm{cut}}$, but the cross section should become insensitive to this choice in the
double-unresolved limit $\tau_1^{\rm{cut}} \rightarrow 0$. 

\section{Results} 

In order to properly define the process of direct photon production it is necessary to apply isolation conditions to the photon. 
Experimentally this reduces the background from unwanted secondary photons, which arise from decays of hadrons. 
An additional source of photons arises from fragmentation of collinear partons.  On the theoretical side, one approach is to model the fragmentation 
process through the introduction of non-perturbative contributions that can be determined by a fit to suitable data.
However, the need to include fragmentation 
functions can be eliminated altogether if the smooth cone isolation, proposed in ref.~\cite{Frixione:1998jh}, is used instead.  This form of isolation is encapsulated by,
\begin{eqnarray}
\sum  E_T^{\rm{had}}(R) < \epsilon_{\gamma} E_T^{\gamma} \left(\frac{1-\cos{R}}{1-\cos{R_0}}\right)^n \quad \forall R < R_0 \,.
\end{eqnarray}
This requirement constrains the sum of the hadronic energy inside a cone of radius $R$, for all separations $R$ that are smaller than
a chosen cone size, $R_0$.  Note that arbitrarily soft radiation will always pass the condition, but collinear $(R \rightarrow 0)$ radiation is forbidden. 
Therefore the contributions from the fragmentation functions are eliminated. This isolation prescription is therefore highly desirable from a theoretical viewpoint. 

Unfortunately, the continuous nature of this isolation prescription cannot be reproduced easily in the experimental setup, in which discrete calorimeter cells are used.
Therefore the smooth cone procedure is not feasible for use in experimental studies.
However, at NLO, it is possible to choose smooth-cone parameters ($\epsilon_\gamma$, $n$ and $R_0$) such that the theoretical prediction using
smooth cone isolation is close to the one obtained using fragmentation functions and isolation conditions that mimic the experimental cuts. 
Such a matching was performed in ref.~\cite{Campbell:2016yrh} for the case of photon pair production. We adopt the same parameters, $n=2$, $\epsilon_\gamma = 0.1$ and $R_0=0.4$, that
were suggested in that study, finding NLO rates that are within $1$-$2$\% of the {\tt{JETPHOX}}~\cite{Catani:2002ny} results quoted in ref.~\cite{Aad:2016xcr}.
\begin{center}
\begin{figure}[ht]
\includegraphics[width=8cm]{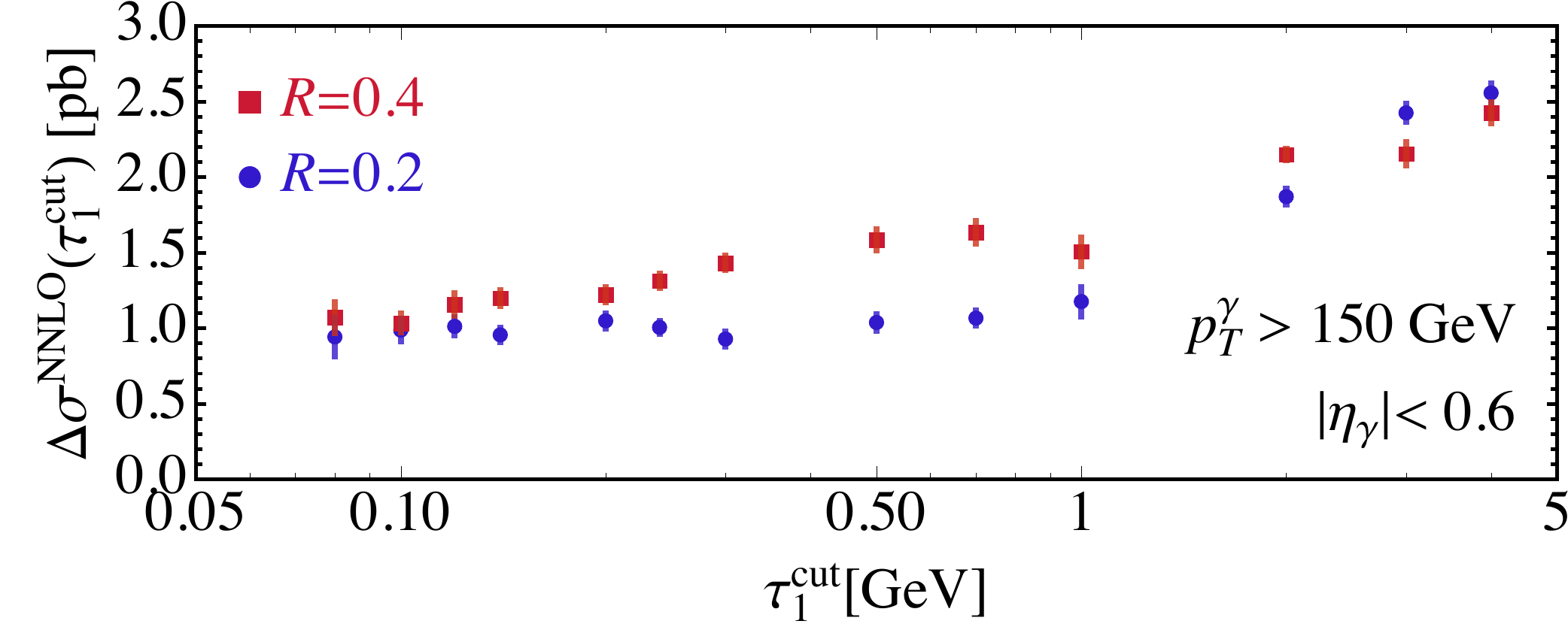}
\caption{The dependence on the NNLO coefficient on the parameter $\tau^{{\rm{cut}}}_1$, and the clustering cone size $R$. Two choices are shown corresponding to $R=0.4$ (red) and $R=0.2$ (blue).}
\label{fig:Rtaudep_dirgam}
\end{figure}
\end{center}

\begin{center}
\begin{figure}[ht]
\includegraphics[width=8cm]{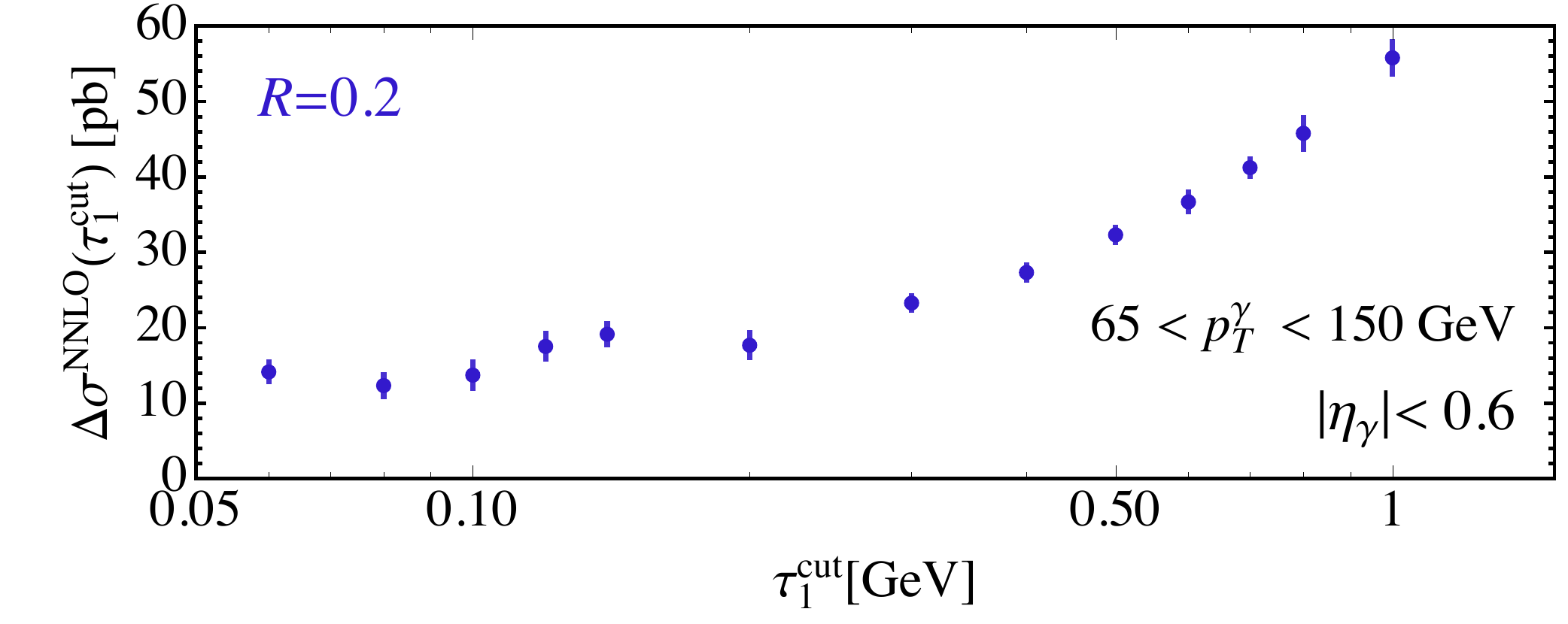}
\caption{The dependence on the NNLO coefficient on the parameter $\tau^{{\rm{cut}}}_1$ for the transverse momentum window $65 < p_T^{\gamma} < 150$ GeV.}
\label{fig:taudep_dirgam65}
\end{figure}
\end{center}
All of the results presented here are for the LHC operating at $\sqrt s=8$~TeV.  The photon is constrained by a simple set of cuts,
\begin{equation}
p_T^\gamma > 65~\mbox{GeV} \,, \qquad |\eta_{\gamma}| < 0.6 \,.
\end{equation}
The theoretical predictions are all obtained using the CT14 NNLO pdf set~\cite{Dulat:2015mca} with renormalization ($\mu_R$) and factorization ($\mu_F$)
scales equal to $p_T^{\gamma}$.  The rate for this process is proportional to the electromagnetic coupling, $\alpha_{em}$. Since our calculation does not include any higher orders
in $\alpha_{em}$ we are free to evaluate this coupling at any scale.   In practice there is a significant difference, of around 7\%, between theoretical
predictions obtained using $\alpha_{em}(0) = 1/137$ and those employing a higher-scale value, $\alpha_{em}(m_Z) = 1/127.9$.  Since we will later
ameliorate our calculation by including the effects of resumming electroweak Sudakov logarithms according to the results of
refs.~\cite{Becher:2013zua,Becher:2015yea} we will adopt the choice therein for all our results, namely the use of $\alpha_{em}(m_Z) = 1/127.9$.
This choice has previously been theoretically motivated in refs.~\cite{Czarnecki:1998tn,Becher:2015yea} and, as we will observe later, it is
supported phenomenologically by an improved description of ATLAS data~\cite{Aad:2016xcr,Schwartz:2016olw}.

In order to validate the method, we first study the dependence of the power corrections on the jet cone size $R$ that is indicated in Fig.~\ref{fig:Rtaudep}.
We compute the NNLO coefficient in the perturbative expansion of the cross-section ($\Delta \sigma^{\rm NNLO}$), for $R=0.2$ and $R=0.4$, for photons
with $p_T^{\gamma} > 150$~GeV.  Our results are shown in Fig.~\ref{fig:Rtaudep_dirgam}.  We observe that for $\tau_1^{\rm{cut}} \gtrsim 0.14$~GeV the
power corrections result in predictions for the NNLO coefficient that are quite different for the two values of $R$. However, for $\tau_1^{\rm{cut}} \lesssim 0.14$~GeV the predictions
tend towards the same result and are in much better agreement. We also note that the smaller cone size has a much flatter dependence on $\tau_1^{\rm{cut}}$.
Although some residual effect from power corrections can be seen for $R=0.2$, the cross section is essentially asymptotic for $\tau_1^{\rm{cut}} \lesssim 0.7$ GeV. 

Given that our calculation is ultimately insensitive to $R$ we can thus choose our value to expedite the onset of asymptotic behavior. We thus choose $R=0.2$ henceforth.
In Figure~\ref{fig:taudep_dirgam65} we present the $\tau^{\rm{cut}}_1$ dependence for the softer region $65 < p_T^{\gamma} < 150$ GeV, which corresponds to the 
softest photons we study in this paper. It is clear that the power corrections are sizable for $\tau^{\rm{cut}}_1 \gtrsim  0.2$ GeV, but that there is little
dependence on $\tau^{\rm{cut}}_1$ in the region $\tau^{\rm{cut}}_1 \le  0.1$ GeV. This is in line
with the expected scaling from the harder $(>  150$ GeV) region we studied previously. For our subsequent comparison with ATLAS data we set
$\tau^{\rm{cut}}_1 = $ \{0.1, 0.2, 0.7\} GeV for the phase space regions $p_T^{\gamma} > $ \{65, 150, 470\} GeV respectively. 

\begin{center}

\begin{figure}[ht]
\includegraphics[width=3.0in]{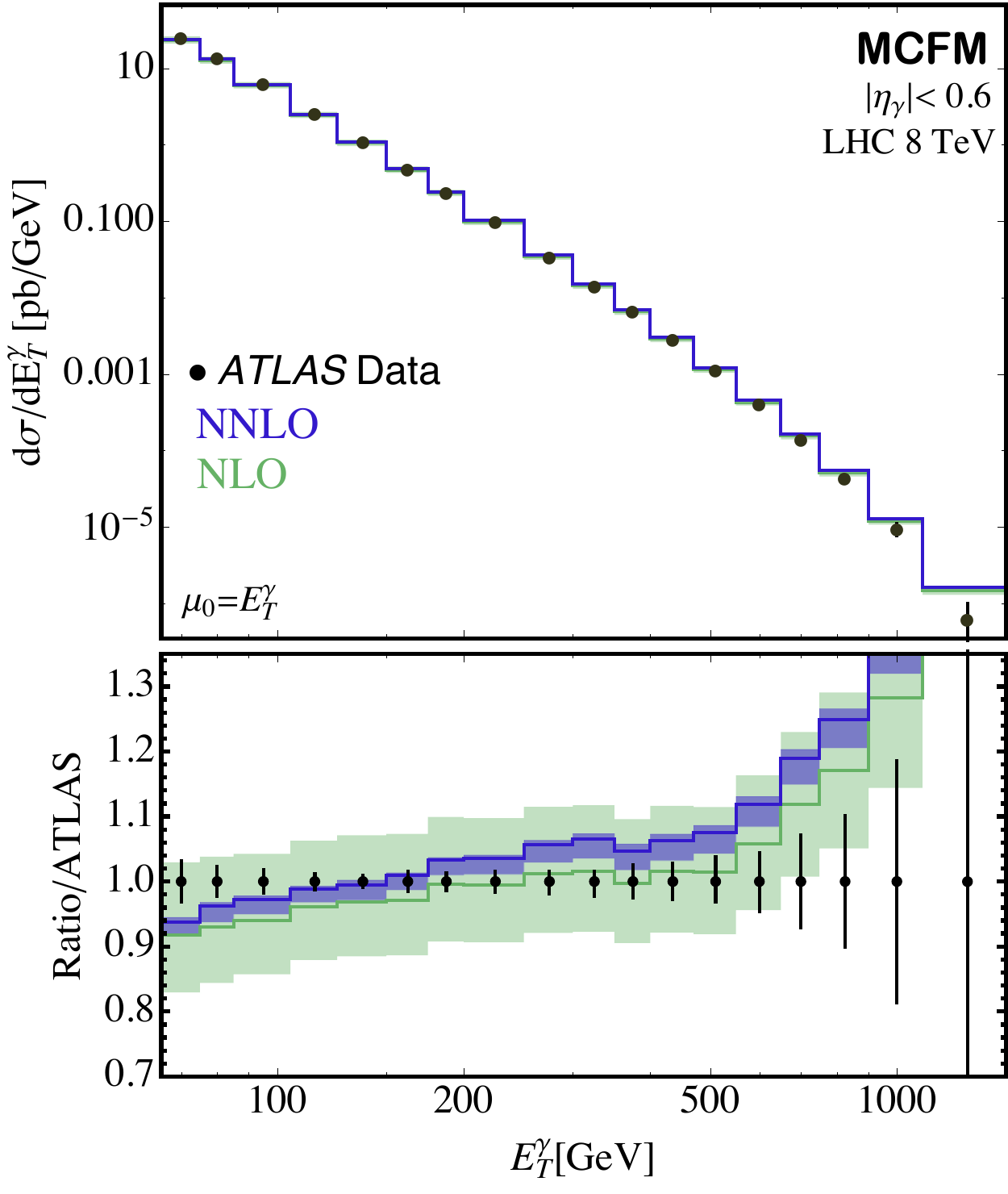}
\caption{A comparison of the MCFM predictions for the transverse momentum of the photon
 to ATLAS 8 TeV data~\cite{Aad:2016xcr}.}
 \label{fig:atlaspty0}
\end{figure}
\end{center}
In Fig.~\ref{fig:atlaspty0} we compare our NNLO (and NLO) predictions from MCFM with 8 TeV ATLAS data~\cite{Aad:2016xcr}. 
The shaded bands represent the scale uncertainty, obtained by considering relative deviations using a six-point scale variation about our central choice:
$\{\mu_R , \mu_F\} =\{ \lambda_1 p_T^{\gamma},  \lambda_2 p_T^{\gamma} \}$ with $\lambda_i \in \{2,1,1/2\}$ and $\lambda_1 \ne \lambda_2^{-1}$. It is clear that the scale
dependence is greatly reduced for the NNLO prediction when compared to NLO. For the central scale choice the NNLO prediction is around 5\% larger
than NLO.   The central scale is close to the maximum of the uncertainty band, with deviations around $+1\%$ and $-4\%$ over much of the range. The tendency of the
theoretical prediction to overestimate the data in the high $p_T$ region is more pronounced when the NNLO correction is included.  This leads to a significant disagreement between
theory and data, far outside the NNLO scale uncertainty band. We note that our larger value of $\alpha_{em}$, results in a much better agreement with data than the lower choice used in~\cite{Aad:2016xcr}
(c.f. also ref.~\cite{Schwartz:2016olw}).

Given the small uncertainty in the NNLO QCD prediction, and the resulting tension with data, it is especially important to investigate the impact
of additional theoretical effects not included in the pure QCD prediction. At high energies it is well-known that the impact of Sudakov effects, arising from the virtual
radiation of heavy electroweak bosons, is important for this process~\cite{Becher:2013zua,Becher:2015yea,Schwartz:2016olw}.  Using a parametrized form that captures the
effect of these leading-logarithmic electroweak corrections to good accuracy~\cite{Becher:2015yea} it is possible for us to also account for these effects.  We thus modify our NNLO prediction
by rescaling it by a factor $[1+\Delta \sigma_V^{ew}(p_T^\gamma)]$, where $\Delta \sigma_V^{ew}(p_T^\gamma)$ is specified in ref.~\cite{Becher:2015yea}.

\begin{center}
\begin{figure}[ht]
\includegraphics[width=3.0in]{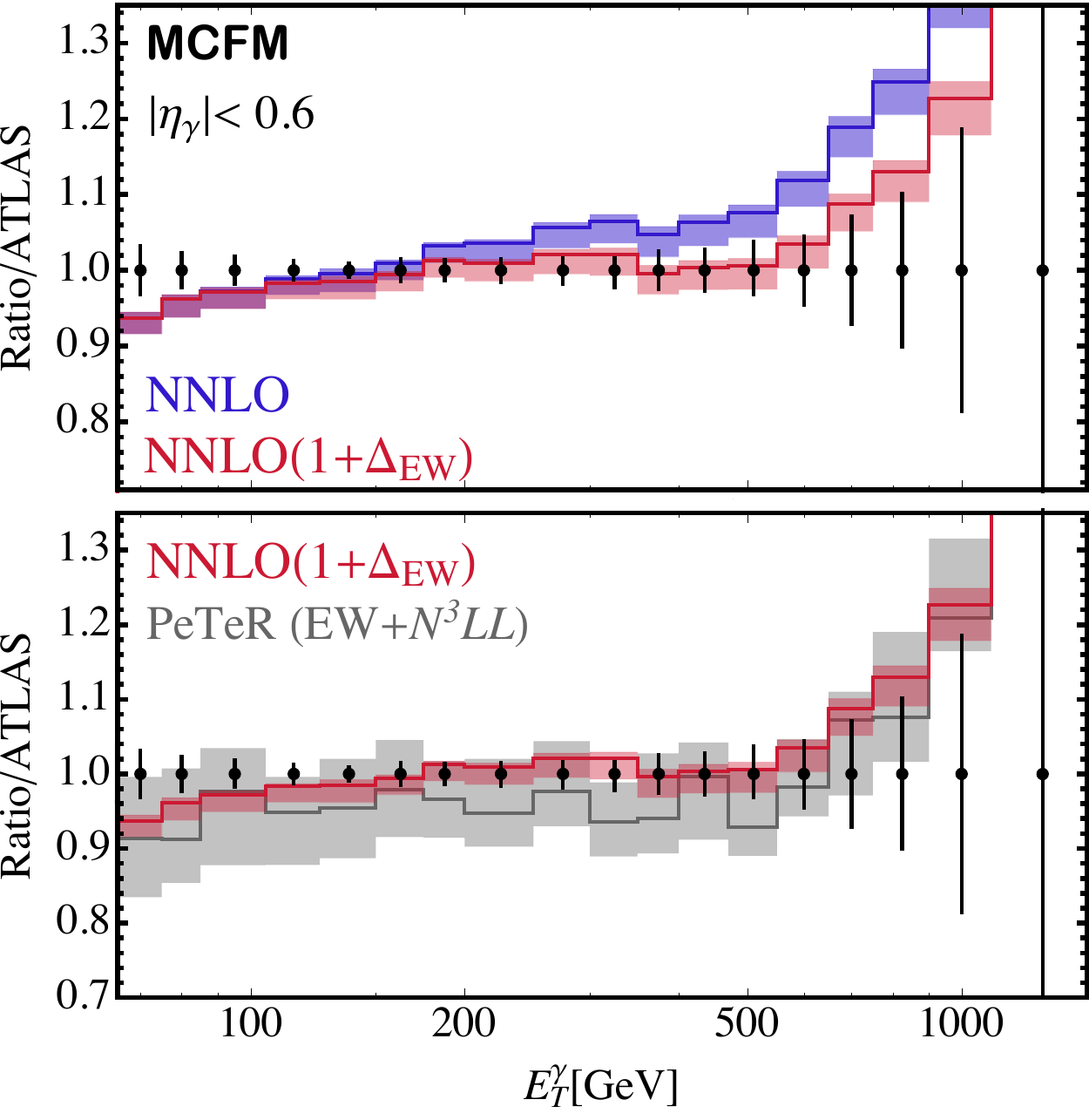}
\caption{Upper: the effect of including electroweak corrections in addition to the NNLO predictions provided by MCFM.
Lower: a comparison of the NNLO+EW prediction of MCFM with the N$^3$LL+EW prediction of {\tt PeTeR}~\cite{Schwartz:2016olw}. }
 \label{fig:EWpetercomp}
\end{figure}
\end{center}

Accounting for both NNLO QCD and electroweak effects in this way provides the improved prediction shown in the top panel of Fig~\ref{fig:EWpetercomp}. This shows a
dramatic improvement in the overall agreement between our theoretical prediction and data after the inclusion of electroweak effects. It is a
remarkable feat that the experimental and theoretical uncertainties are now under such good control that the inclusion of the electroweak corrections
becomes mandatory to ensure agreement between theory and data at energies as low as a few hundred GeV.  To indicate the level of improvement that the NNLO QCD corrections
provide, the lower panel shows a comparison of our best prediction and the previous most accurate calculation presented in ref.~\cite{Schwartz:2016olw}.
The result of ref.~\cite{Schwartz:2016olw}, obtained using the {\tt{PeTeR}} code, accounts for threshold resummation to N$^{3}$LL accuracy and also includes the same electroweak effects.
It is clear from the figure that the central values for the two predictions are similar.  However the scale uncertainty in the NNLO
calculation is smaller, by around a factor of three, than the equivalent uncertainty obtained using {\tt{PeTeR}}. 

\section{Conclusions}
\label{sec:conc}

We have presented a calculation of direct photon production at NNLO accuracy obtained using the $N$-jettiness slicing approach. We compared our prediction to 
ATLAS 8 TeV data for $p_T^{\gamma} > 65$ GeV and $|\eta_{\gamma}| < 0.6$. We found that by combining the NNLO QCD calculation with EW effects our calculation
describes the data very well. Our results represent a significant improvement compared to previous theoretical predictions. The future study of this process, over a wider phase space and at larger
center of mass energies, presents an exciting opportunity for precision QCD at colliders. In particular, the calculation of ratios of photon momenta for different rapidity regions has interesting potential. The ratios have the advantage of cancelling the leading 
dependence on $\alpha_{em}$ and simultaneously the experimental luminosity uncertainty.  Theoretical predictions for these ratios at NNLO could be used to constrain pdfs, provided
that remaining theoretical uncertainties, such as those related to isolation, are fully understood. We leave such a detailed phenomenological study to a future publication. 

\section*{Acknowledgements}

We are extremely grateful to Radja Boughezal, Xiaohui Liu and Frank Petriello for providing us with a 
numerical fit to their calculation of the soft $1$-jettiness function that was reported in ref.~\cite{Boughezal:2015eha}.
Support provided by the Center for Computational Research at the University at Buffalo.
CW wishes to thank Sal Rappoccio for providing access to additional CCR computer cores.
CW is supported by the National Science Foundation through award number PHY-1619877.
The research of JMC is supported by the US DOE under contract DE-AC02-07CH11359.


\bibliography{dirgam} 

\end{document}